# Strain tuned magnetotransport of $J_{\text{eff}}$=1/2 antiferromagnetic Sr$_2$IrO$_4$ thin films

N. Hu[1#], Y. K. Weng[2,3#], K. Chen[1], B. You[4], Y. Liu[5], Y. T. Chang[4], R. Xiong[5], S. Dong[3], and C. L. Lu[4*]

[1] *School of Science and Hubei Collaborative Innovation Center for High-Efficiency Utilization of Solar Energy, Hubei University of Technology, Wuhan 430068, China*

[2] *School of Science, Nanjing University of Posts and Telecommunications, Nanjing 210023, China*

[3] *School of Physics, Southeast University, Nanjing 211189, China*

[4] *School of Physics & Wuhan National High Magnetic Field Center, Huazhong University of Science and Technology, Wuhan 430074, China*

[5] *School of Physics and Technology, and the Key Laboratory of Artificial Micro/Nano structures of Ministry of Education, Wuhan University, Wuhan 430072, China*

---

[#] These authors contributed equally to this work.
[*] Email: cllu@hust.edu.cn

**Abstract:**

In this work, we report observation of strain effect on physical properties of $Sr_2IrO_4$ thin films grown on $SrTiO_3$ (001) and $LaAlO_3$ (001) substrates. It is found that the film on $LaAlO_3$ with compressive strain has a lower antiferromagnetic transition temperature ($T_N$~210 K) than the film on $SrTiO_3$ ($T_N$~230 K) with tensile strain, which is probably caused by modified interlayer coupling. Interestingly, magnetoresistance due to pseudospin-flip of the film on $LaAlO_3$ is much larger than that of tensile-strained film on $SrTiO_3$, and robust anisotropic magnetoresistance is observed in the former, but *H*-driven reversal behavior is seen in the latter. By performing first principles calculations, it is revealed that epitaxial strain plays an efficient role in tuning the canting angle of $J_{eff}$=1/2 moments and thus net moment at every $IrO_2$ layer, responsible for the difference in magnetoresistance between the films. The reversal of anisotropic magnetoresistance in the thin film on $SrTiO_3$ can be ascribed to stabilization of a metastable stable with smaller bandgap as the $J_{eff}$=1/2 moments are aligned along the diagonal of basal plane by *H*. However, theoretical calculations reveal much higher magnetocrystalline anisotropy energy in the film on $LaAlO_3$. This causes difficulties to drive the $J_{eff}$=1/2 moments to reach the diagonal and thereby the metastable state, explaining the distinct anisotropic magnetoresistance between two samples in a qualitative sense. Our findings indicate that strain can be a highly efficient mean to engineer the functionalities of $J_{eff}$=1/2 antiferromagnet $Sr_2IrO_4$.

## I. Introduction

The layered perovskite $Sr_2IrO_4$ (SIO), hosting a novel $J_{eff}=1/2$ Mott state and a quasi-two-dimensional (2D) square-lattice antiferromagnetic (AFM) phase, provides a fertile ground for exploring emergent physics that arises from the interplay of various fundamental interactions such as strong spin-orbit interaction (~0.5 eV), electron correlation, and crystal field [1-3]. The $J_{eff}=1/2$ Mott state as a profound manifestation of the strong spin-orbit coupling was first experimentally identified in $Sr_2IrO_4$ in 2008 [4], and its main picture can be described as following. Due to the octahedral crystal filed, the 5$d$ orbitals of $Ir^{4+}$ are split into the twofold $e_g$ orbitals and threefold $t_{2g}$ orbitals. Strong spin-orbit coupling further causes orbital splitting, and thus induces two sub-bands, i.e. half-filled $J_{eff}=1/2$ band and fully occupied $J_{eff}=3/2$ band. The $J_{eff}=1/2$ band is narrow enough, and a moderate coulomb repulsion is sufficient to open a charge gap of ~0.3 eV in $Sr_2IrO_4$ [2]. Interestingly, such a spin-orbit coupling assisted Mott phase was found to closely associate with the lattice degree of freedom, i.e. the band-gap was predicted to strongly depend on the Ir-O-Ir bond angle $\phi$ [5].

Another defining character of $Sr_2IrO_4$ is the $J_{eff}=1/2$ AFM phase on a quasi-2D square lattice, which highly resembles $La_2CuO_4$, a parent compound of the high temperature ($T$) superconductors. As sketched in Fig. 1, the $J_{eff}=1/2$ magnetic moments (entangling both the spin and orbital moments) lie in-plane and possess obvious canting relative to the crystalline axes [6,7]. Impressively, it has been demonstrated that the canting angle $\alpha$ of $J_{eff}=1/2$ moments closely tracks the rotation angle $\varphi$ of $IrO_6$ octahedra, i.e. the so-called pseudospin-lattice locking effect $\alpha \sim \varphi$ [8,9]. This is seldom seen in traditional 3$d$ transition metal oxides. Because of the remarkable orbital magnetism, the $J_{eff}=1/2$ moments are sensitive to crystalline symmetry, which is responsible for the bond-dependent magnetism in honeycomb lattice such as $Na_2IrO_3$, but Heisenberg interactions in square lattice such as SIO [1]. And, the magnetic easy axis of SIO notably points away from the Ir-O-Ir bond. Recently, it has been clarified that pseudospin-lattice coupling is crucial for the in-plane magnetic anisotropy [10,11].

These previous studies have illustrated the close relationship between the lattice degree of freedom and the electronic structures in SIO. Inspired by this mechanism, considerable attentions have been focused on the modulation of physical properties of $J_{eff}$ states (not only the $J_{eff}=1/2$ but also the $J_{eff}=3/2$ one) by tuning the lattice distortion using strain, and fascinating phenomena have been found or predicted in the past few years [12-16]. For instance, theoretical calculations revealed that the $J_{eff}$ states, AFM configuration, and even the $J_{eff}$ moment itself (i.e. the ratio of spin and orbital moments) can be modified drastically by varying lattice parameters such as $c/a$ ratio [14-16]. Some of these predictions have been observed in experiments recently [17-19].

However, only few examples of strain effect on magnetoresistance (MR) exist in literature, which should be a natural and attractive continuation of this topic. Because of the canting of $J_{eff}=1/2$ moments, net moments arise in $IrO_2$ layers, which are ordered in an AFM manner along the $c$-axis without showing macroscopic magnetization ($M$). By applying a magnetic field $H$, a spin-flip transition is triggered when $H$ is large enough, and the net moments of $IrO_2$ layers are switched to align ferromagnetically [6]. This process is accompanied by a sudden reduction in resistivity ($\rho$) [20], reminiscent of the classic giant magnetoresistance (GMR). Nevertheless, here it takes place in an atomic scale and in an AFM phase. Miao *et al.* revealed that the GMR-like effect became dispersive due to anisotropic tensile strain in SIO thin films [21]. Differently, Zhang *et al.* found that the MR-drop can be even sharper when in-situ anisotropic strain is applied in bulk SIO [22]. Beyond these, SIO possesses more alluring MR behaviors such as the remarkable anisotropic magnetoresistance (AMR) [23,24]. In particular, the magnitude of AMR of SIO is far larger than many other antiferromagnets such as the AFM alloys containing $5d$ elements. This is factually compelling, since obtaining large AMR has been a core issue in the field of AFM spintronics [25,26]. In our previous work, it was found that the fourfold AMR can be reversed once $H$ is sufficiently high, which was ascribed to bandgap engineering due to rotation of $J_{eff}=1/2$ moments [27], rarely seen in conventional magnetic materials. Motivated by the versatile MR phenomena and the very active role of lattice in tuning properties in SIO, it is of high interest and actuality to comprehensively investigate strain effect on magneto-transport of SIO, which would in turn be helpful for understanding the essence of $5d$ physics.

According to previous theoretical calculations, compressive and tensile strain has asymmetric effects on the $J_{eff}=1/2$ state and AFM phase, i.e. compressive (tensile) strain tends to drive $Sr_2IrO_4$ toward (away) the ideal $J_{eff}=1/2$ point [14]. Therefore, in the present work, SIO epitaxial thin films were grown on (001) $SrTiO_3$ (STO) and (001) $LaAlO_3$ (LAO) substrates to have tensile and compressive strain, respectively. Extensive magneto-transport measurements of the thin films were carried out. We find that the SIO/LAO films with compressive strain show stronger ferromagnetism and much larger MR than the tensile-strained SIO/STO films. Fourfold AMR related to magnetocrystalline anisotropy was observed for both types of films. However, the crystal AMR of SIO/STO can be reversed as *H* is above a critical value, but it behaves quite robust against *H* in SIO/LAO. By performing first principles calculations, it is found that the Ir-O-Ir bonds tend to be bended (stretched) by compressive (tensile) strain, and the canting angle *α* of $J_{eff}$=1/2 moments get to be increased (reduced) concurrently, which explains the enhanced ferromagnetism and sequential MR. Meanwhile, it is revealed that a hidden metastable state with smaller bandgap may be approached as $J_{eff}$=1/2 moments are aligned along the Ir-O-Ir bond, i.e. diagonal of basal plane. However, in SIO/LAO films, larger magnetic anisotropy is obtained due to the compressive strain as compared with SIO/STO, which makes it harder to access the metastable state, responsible for the sturdy AMR.

## II. Experiments

$Sr_2IrO_4$ thin films of ~40 nm were simultaneously grown on STO (001) and LAO (001) substrates using pulsed laser deposition system equipped with in-situ reflection high energy electron diffraction (RHEED). The growth parameters were carefully optimized, and the growth rate was calibrated using RHEED and transmission electron microscopy. More details of the thin film growth can be found in the previous studies [28,29]. In order to examine the crystalline structure of the films, X-ray diffraction (XRD) measurements such as regular theta-2theta scan, rocking curve, and reciprocal space mapping (RSM) were carried out using a Philips X'Pert diffractometer. Magnetization characterizations, including both *M*(*T*) and *M*(*H*), were performed using a superconducting quantum interference device (Quantum Design) with sensitivity of ~$1\times10^{-8}$ emu at *H*=0 T and ~$8\times10^{-8}$ emu at *H*=7 T. *M*(*T*) curves of both field cooling (FC) and zero field cooling (ZFC) sequences were measured with *H* = 0.1 T. Standard four-probe method was used for electric transport measurements with electric current *I*//[100] in a physical property measurement system. With regard to the AMR measurements, *I* was applied always along the [100] direction, and *H* was rotated within basal plane of the films. The AMR curves were collected at various *T* below $T_N$ and *H* ranging from 0 to 9 T (the maximum value allowed by our transport measurement set-up).

Density functional theory (DFT) calculations using Vienna *ab initio* Simulation Package (VASP) were performed [30,31]. The Hubbard repulsion $U_{eff}$ = 3 eV was imposed on Ir's 5d orbitals and spin orbit effect was considered with noncollinear spins. The plane-wave cutoff was 550 eV and the 9 × 9 × 2 Monkhorst–Pack k-points mesh was centered at Γ points. To simulate the strain effects, the in-plane lattice constants of films were fixed to be the same as the substrates. The $J_{\text{eff}}=1/2$ moments were initialized along particular axes (without canting), and the magnetic canting was obtained via self-consistent calculations.

**III. Results and discussions**

Figures 2(a) and 2(b) show theta-2theta scans of the films. Both samples are pure phase, and show single *c*-axis orientation. Rocking curves around the film SIO (004) (red and blue circles), and corresponding substrates STO (001) and LAO (001) reflections (black lines) are shown in Fig. 2(c) and 2(d), respectively. The full width at half maximum (FWHM) of film is close to that of underlying substrate in every sample, indicating the good crystallinity of the thin films. Our previous scanning transmission electron microscopy characterization evidenced very good atom layers in the films [29]. According to the RSM data shown in Fig. 2(e), SIO thin films is coherent to STO substrate, and the in-plane and out-of-plane lattice parameters are estimated to $a/\sqrt{2}=3.905$ Å and $c/2=12.848$ Å, respectively, giving rise to slight tensile strain of ~ 0.4%. For SIO/LAO, the derived lattice parameters are $a/\sqrt{2}=3.858$ Å and $c/2=12.908$ Å, corresponding to compressive strain of ~ - 0.8%, smaller than the ideal value of -2.5% based on the lattice parameters of bulk counterparts [2]. This is consistent with the RSM result which shows strain relaxation of the film. For convenience, here a pseudo-tetragonal lattice expression is used, which has a 45° in-plane rotation with respect to the tetragonal lattice of SIO. For instance, the *a*-axis of SIO is now along the [110] direction.

Because of strain, the two samples show different magnetic properties. As shown in Fig. 3(a), the in-plane *M*(*T*) curves of both FC and ZFC sequences of SIO/STO exhibit abrupt enhancement at $T_N$~230 K, arising from development of weak ferromagnetism due to $J_{\text{eff}}=1/2$ moment canting. This is comparable with the value of bulk SIO [32]. Clearly, *M* of out-of-plane is far smaller than that of in-plane, evidencing the quasi-2D nature of the AFM phase with magnetic moments aligned in-plane. For SIO/LAO, while the striking magnetic anisotropy is maintained, the weak ferromagnetic transition is reduced to $T_N$ ~210 K. As revealed by previous studies [1,33], while the intralayer coupling (~60 meV) is far stronger than the weak interlayer interaction (~1 μeV), the latter plays a crucial role in stabilizing the long range AFM order in SIO [33,34], akin to the case of La$_2$CuO$_4$. The strength of interlayer coupling depends on the distance of IrO$_2$ layers, which is tunable by the $c/a$ ratio. For instance, reducing $c/a$ can strengthen the interlayer coupling, and thus move $T_N$ toward higher $T$, as confirmed by recent experiments [32]. This agrees well with our observations that SIO/STO has smaller $c/a$ and thereby a higher $T_N$ than SIO/LAO.

To capture more details of the canted AFM phase, magnetization as a function of $H$ were measured at $T=10$ K (far below $T_N$) for the two samples. First, $M(H)$ curves of both samples (with substrates) and pure substrates were measured separately. Substrates that used for diamagnetic property measurements and thin film growth were cut off from the same bigger piece, in order to precisely determine the magnetic properties of thin films. Second, $M(H)$ curves of pure thin films were obtained after subtracting the diamagnetic contribution of substrates, shown in Fig. 3(c). The uncertainty of $M(H)$ such as at high field range may arise from the small net magnetization of Sr$_2$IrO$_4$, and the correction process. Nevertheless, the weak ferromagnetic phase due to canting of the $J_{eff}=1/2$ magnetic moments can still be well characterized, benefitting from the high sensitivity of instrument. In Fig. 3(c), it is seen that SIO/LAO has much larger saturated $M_a$ than SIO/STO. Since macroscopic $M_a$ of the weak ferromagnetic phase simply arises from net moments of IrO$_2$ layers due to canting of $J_{eff}=1/2$ moments, i.e. $M_a=M_{Ir}\cdot\sin\alpha \approx M_{Ir}\cdot\sin\varphi$, where $M_{Ir}$ is the Ir moment, the larger $M_a$ indicates enlarged canting angle $\alpha$ in SIO/LAO as compared with SIO/STO. As schematically shown in Fig. 3(d), epitaxial strain effectively exerts biaxial pressure to the in-plane Ir-O-Ir bond network, and compressive (tensile) strain tends to bend (expand) the Ir-O-Ir bond angle, resulting in larger (smaller) rotation angle $\varphi$ of IrO6 octahedra. As a consequence, $\alpha$ is modified coherently, because of the locking effect $\alpha \sim \varphi$. This is supported by further first principles calculations shown below.

Based on the spin-valve-like configuration of the quasi-2D AFM phase, enhancing net moment of IrO$_2$ layers may lead to larger MR. With this motivation, comprehensive MR measurements have been carried out. As expected, insulating transport behavior is evidenced in both SIO/STO and SIO/LAO thin films (Fig. 4(a)), arising from the $J_{eff}=1/2$ Mott state. Fig. 4(b) shows measured MR at various temperatures for two samples. MR-drop associated with the spin-flip transition can be seen in both samples. However, SIO/LAO possesses more obvious MR-drop and much larger MR than SIO/STO. For instance, at $T=30$ K and $H=9$ T, derived MR of SIO/LAO is ~10%, which is about one order of magnitude larger than that of SIO/STO (MR~1%). In addition, critical field of the MR-drop is evidently higher for SIO/LAO than SIO/STO, consistent with the rigorously expanded $M(H)$ loop in SIO/LAO (Fig. 3(c)). These facts evidence higher energy difference between the AFM and the induced weak ferromagnetic states in the compressively strained SIO/LAO thin film.

Another interesting phenomenon is that SIO/STO shows an abnormal intercross at $H$~3 T between the MR curves with $H//[100]$ and $H//[110]$, shown in Fig. 4(c). During the MR measurements, electric current $I$ is always applied along the [100] direction, while $H$ is applied along either the [100] or the [110] directions. As mentioned above, the magnetic easy axis of SIO, i.e. the $a$-axis, is along the [110] direction. In this sense, $J_{\text{eff}}=1/2$ moments are usually easier to be aligned as $H$ is applied along the [110] direction than $H//[100]$, giving rise to stronger suppression of magnetic scattering and thus larger MR. This normal situation is seen at low field region such as $H$<3 T. However, at $H$>3 T, MR of $H//[100]$ (i.e. the hard axis) becomes even larger than that of $H//[110]$ (i.e. the easy axis), indicating exotic mechanism is involved in dictating magneto-transport of SIO/STO. Differently, for SIO/LAO, MR of $H//[110]$ remains larger than that of $H//[100]$ till $H$=9 T, which follows conventional perception based on magnetic anisotropy.

In order to gain further insight into strain tuned MR of the thin films, AMR curves at various $H$ and $T$ were measured for two samples. During AMR measurements, electric current $I$ was applied along the [100] direction, $H$ was rotated within the basal plane, and $\Phi$ is defined as the angle between $H$ and $I$. As shown in Fig. 5(a), the low-$H$ (i.e. $H$<0.5 T) AMR is more like twofold, while clear peak-splitting arises around $\Phi$~45° and $\Phi$~225°. At $H$=0.5 T, the AMR curve (thick olive line) shows four minima appearing at $\Phi \sim 53°+n\pi/2$ ($n$=0,1,2,3), which matches with the magnetic easy axes of $Sr_2IrO_4$. Such AMR effect showing minima at magnetic easy axes can be seen until $H$~4 T (thick red line) which is above the anisotropy field ~1.5 T (Fig. 3(c)). Therefore, the AMR should be majorly fourfold due to the magnetocrystalline anisotropy, consistent with previous studies [35,36]. Nevertheless, at 0.5 T<$H$<4 T, peak positions of the fourfold AMR show slight shift upon increasing $H$, which is probably due to the competition between Zeeman energy and magnetic anisotropy. At $H$=7 T, a new set of peaks start to arise in the places where valleys were observed, and the original peaks of the low-$H$ fourfold AMR are suppressed simultaneously. This is getting more striking at $H$=9 T, i.e. the AMR curve is reversed. Since the original peaks/valleys have not been completely suppressed, the AMR at $H$=9 T displays a mixed behavior, i.e. various oscillation components coexisted. At $T$=100 K and $H$=9 T shown in Fig. 5(b), the AMR-reversal can be more clearly seen. However, for SIO/LAO, the fourfold AMR looks quite robust against $H$, and AMR-minima remain unmoved until $H$=9 T, i.e. the AMR minima are kept at the [110] and equivalent directions (magnetic easy axes), shown in Fig. 5(c). The AMR-minima positions as a function of $H$ of both samples are shown in Figs. 5(d) and (e). For SIO/STO, the AMR-reversal behavior can be clearly identified, and the estimated critical field (indicated by arrows) is reduced as $T$ is increased. For SIO/LAO, the AMR-minima evolve with $H$ steadily without showing any apparent shift. The striking difference in AMR between two samples suggests the significant role of strain in tuning magnetic anisotropy of SIO.

To understand underlying physics of the strain effects on magneto-transport of the films, we performed DFT calculations using VASP. In our previous work, it was found that the AMR-reversal in SIO/STO thin films can be ascribed to bandgap engineering due to rotation of $J_{eff}$=1/2 moments. For instance, an evidently smaller bandgap is expected as $J_{eff}$=1/2 moments are aligned along the [100] direction (i.e. diagonal of basal plane), in comparison with the ground state where $J_{eff}$=1/2 moments point to the [110] direction (i.e. the $a$-axis) [27]. In the present work, we focus on the strain effects on magneto-transport of SIO thin films with different strain state, i.e. SIO/STO with tensile strain of ~0.4% and SIO/LAO with compressive strain of ~ -0.8%.

The calculations were first performed for unstrained case, i.e. bulk SIO. The quasi-2D AFM ground state with magnetic easy axis along the [110] direction is confirmed. Besides, some key parameters such as canting angle ~12.5°, local magnetic moment of Ir ~0.498 $\mu_B$, orbital moment $\mu_L$~0.341 $\mu_B$, spin moment $\mu_S$~0.157 $\mu_B$, and the ratio of $\mu_L/\mu_S$~2.2 agree well with previous experimental and theoretical results [7,37]. Then, we turn to calculate strain effect of the films. For both SIO/STO and SIO/LAO, the magnetic easy axis remains at [110], consistent with the fourfold AMR symmetry at low field. Slight variation is found for the magnitude of $J_{eff}$=1/2 moment, resulting from modest modulation of $\mu_L$ and $\mu_L$. The $\mu_L/\mu_S$ ratio is increased to ~2.6 for SIO/STO, but decreased to ~2.1 for SIO/LAO, in agreement with the results calculated using different methods [14]. However, the canting of $J_{eff}$=1/2 moment looks quite sensitive to strain, and two samples show distinct $\alpha$, i.e. $\alpha$~7.7° for SIO/STO and $\alpha$~14.5° for SIO/LAO. Since $M_a$ is determined by $\alpha$ as mentioned above, the larger $\alpha$ evidences larger $M_a$ in SIO/LAO, in comparison with SIO/STO. As proposed by Haney *et al.*, GMR can also occur in an AFM analogue of spin-valve, which was primarily due to the interface spin polarization [38]. However, such AFM GMR requires very high quality interface, and remains elusive in experiments to date due to the challenge of fabricating realistic device with perfect epitaxy. Here SIO hosts a natural AFM spin-valve-like structure, i.e. stacking of AFM $IrO_2$ layers with net moment, which provides a rare platform to study the AFM-GMR. Therefore, the observation of strikingly enhanced MR associated with increased $M_a$ in SIO/LAO represents an experimental evidence of Haney's theory.

As listed in Table I, the magnetocrystalline anisotropy energy $\Delta E$ of $J_{eff}$=1/2 moment aligned along the [110] direction and the [100] direction also shows strong dependence on the strain. For instance, the calculated $\Delta E$ is ~1.28 meV/u.c. for SIO/LAO, but an obviously smaller $\Delta E$ ~1.01 meV/u.c. is revealed for SIO/STO. Similar anisotropic energy of $\Delta E$ ~1.52 meV/u.c. was also reported in $Ba_2IrO_4$ [39]. If only the competition due to Zeeman energy is considered, a moderate $H$ (~10 T at $T$=0 K) exerting on $M_{Ir}$ ~0.3 $\mu_B$ would be sufficient to overcome the anisotropy energy barrier in SIO/STO, and thus drive the $J_{eff}$=1/2 moments to along the [100] direction where the metastable state with smaller bandgap is approached. As a consequence, the AMR is reversed by $H$ in SIO/STO. However, much larger magnetocrystalline anisotropy energy is found in SIO/LAO, because of the compressive strain. In this sense, a much higher external $H$ would be required to drive the $J_{eff}$=1/2 moments to approach the [100] direction and thereby the metastable state in SIO/LAO. This could qualitatively explain the absence of AMR-reversal in SIO/LAO in this work.

At last, we briefly discuss possible experimental evidence of the AMR-reversal in SIO/STO thin films. First, one may expect variation in magnetic anisotropy, corresponding to the AMR-reversal. However, $M(H)$ curves of $H$//[100] and $H$//[110] show negligibly small difference at high field range where the AMR-reversal happens (not shown here). Two possible reasons may be proposed: One is the very small net magnetization due to canting of the $J_{eff}$=1/2 magnetic moments. The other one is that the anisotropic magnetization is small indeed, as revealed by previous studies [10, 40]. Second, such band-gap modulation with different moment orientation has been proposed in many other magnetic materials, and characterizations using angle-resolved photoemission spectroscopy and scanning tunneling microscopy equipped with vector magnetic field would be recommended to visualize the band-engineering [41,42].

## IV. Conclusion

In conclusion, strain effect on the magneto-transport of $Sr_2IrO_4$ thin films grown on STO (001) and LAO (001) substrates has been studied by performing extensive structural, magnetization, and MR measurements. It is found that pseudospin-flip induced MR of SIO/LAO is far larger than SIO/STO, i.e. the difference can be one order of magnitude at $T$=30 K. Fourfold AMR due to magnetocrystalline anisotropy is evidenced for both films. However, SIO/STO exhibits AMR-reversal related to bandgap engineering, but SIO/LAO hosts robust AMR effects against $H$. Our first principles calculations reveal that compressive (tensile) strain can increase (decrease) the canting angle of $J_{eff}$=1/2 moments and thus net moment at every $IrO_2$ layer, responsible for the difference in MR. In the meanwhile, an obvious higher magnetocrystalline anisotropy energy is obtained for SIO/LAO, which hinders the $J_{eff}$=1/2 moments to be move to the [100] direction and thereby the metastable state with smaller bandgap. These findings indicate that epitaxial strain is efficient in tuning magnetotransport of $Sr_2IrO_4$.

**Acknowledgements:** This work is supported by the National Nature Science Foundation of China (Grant No. 12174128, No. 11834002, No. 11804168, and No. 12074291), Hubei Province Natural Science Foundation of China (Grant No. 2020CFA083).

**Figure captions:**

Figure 1. Sketch of crystalline and magnetic structure of $Sr_2IrO_4$. Magnetic moments of Ir ions (red balls) are indicated by black arrows, and net moments of $IrO_2$ layers due to canting are indicated by orange arrows. The pseudospin-lattice locking effect is illustrated at the right side.

Figure 2. XRD patterns for (a) $Sr_2IrO_4$/$SrTiO_3$ (001), and (b) $Sr_2IrO_4$/$LaAlO_3$ (001) thin films. (c) and (d) show rocking curves of two films. Rocking curves of substrates are also shown together (black lines) (e) Reciprocal space mappings around the (109) reflection of the films. (f) TEM images of the SIO/LAO thin film. In the STEM image (right), Sr and Ir atoms are indicated by green and red dots, respectively. The $c$ value is derived to be ~12.92 Å.

Figure 3. $T$-dependence of magnetization for (a) SIO/STO, and (b) SIO/LAO thin films. (c) $M(H)$ curves measured with $H$//[110] at $T=10$ K for two samples. (d) Schematic drawing of strain effect on the canting of $J_{eff}=1/2$ moment.

Figure 4. (a) Resistivity as a function of $T$ for two samples. (b) Measured magnetoresistance at various temperatures for the samples. Magnetoresistance measured with $H$ applied along different directions, i.e. [100] and [110], at $T=30$ K for (c) SIO/STO, and (d) SIO/LAO.

Figure 5. (a) and (b) show anisotropic magnetoresistance measured at various $H$ and $T$ for SIO/STO thin film. For a better view, the AMR curves in (a) have been shifted vertically. (c) AMR curves measured under various magnetic fields at $T=100$ K for SIO/LAO thin film. (d) and (e) present minima positions of the AMR curves as a function of $H$ for SIO/STO and SIO/LAO, respectively.

**Table I.** The energy difference for a minimal unit cell (eight formula units), canting angle of $J_{eff}=1/2$ moments, local spin moment ($\mu_S$) within the default Wigner-Seitz sphere, and orbital moment ($\mu_L$) for each Ir of $Sr_2IrO_4$.

|  |  | 110 | 100 |
|---|---|---|---|
| Unstrained SIO | Energy (meV) | 0 | 1.24 |
|  | Canting angle (°) | 12.5 |  |
|  | $\mu_S$ ($\mu_B$) | 0.157 | 0.155 |
|  | $\mu_L$ ($\mu_B$) | 0.341 | 0.341 |
| SIO/STO (001) | Energy (meV) | 0 | 1.01 |
|  | Canting angle (°) | 7.7 |  |
|  | $\mu_S$ ($\mu_B$) | 0.134 | 0.131 |
|  | $\mu_L$ ($\mu_B$) | 0.343 | 0.343 |
| SIO/LAO (001) | Energy (meV) | 0 | 1.28 |
|  | Canting angle (°) | 14.5 |  |
|  | $\mu_S$ ($\mu_B$) | 0.163 | 0.161 |
|  | $\mu_L$ ($\mu_B$) | 0.340 | 0.340 |

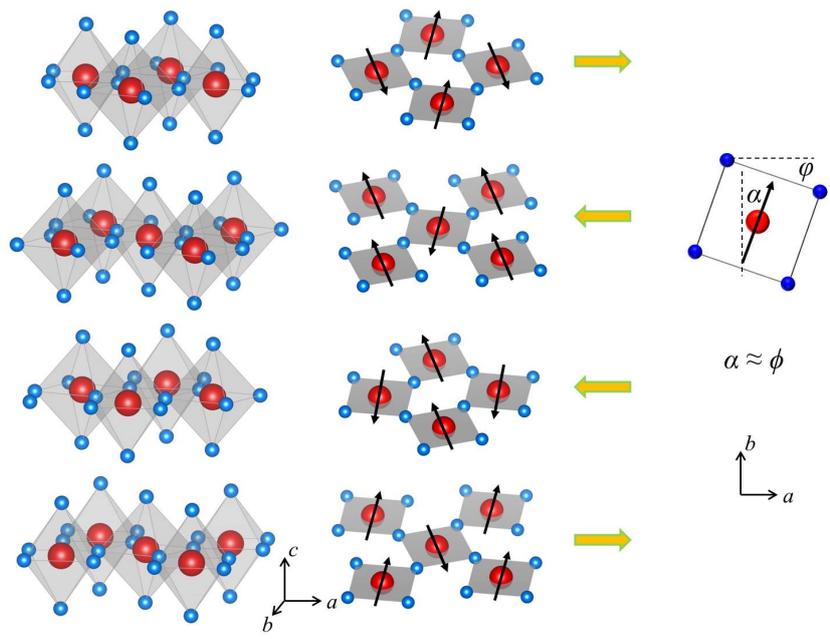

Figure 1

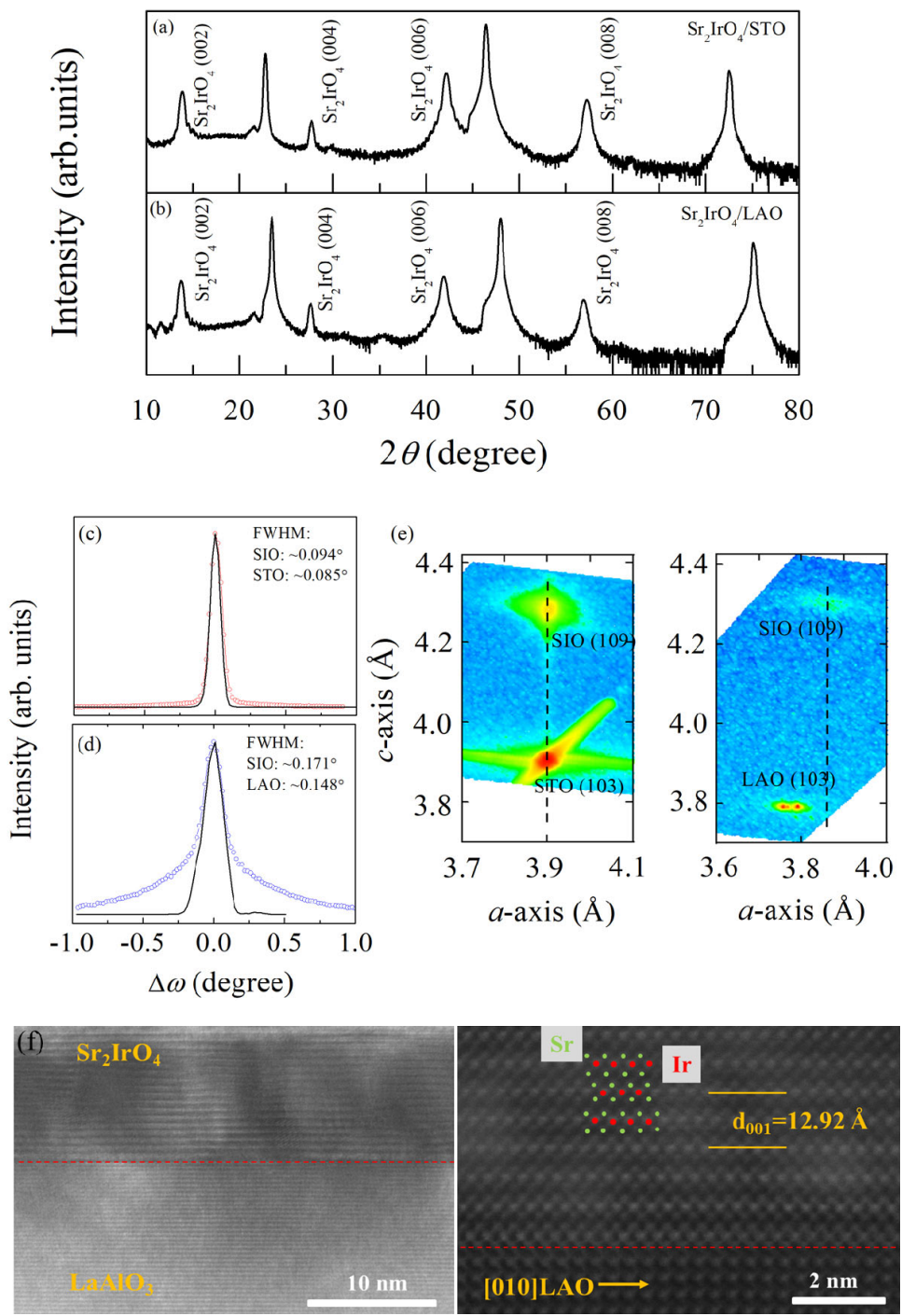

Figure 2

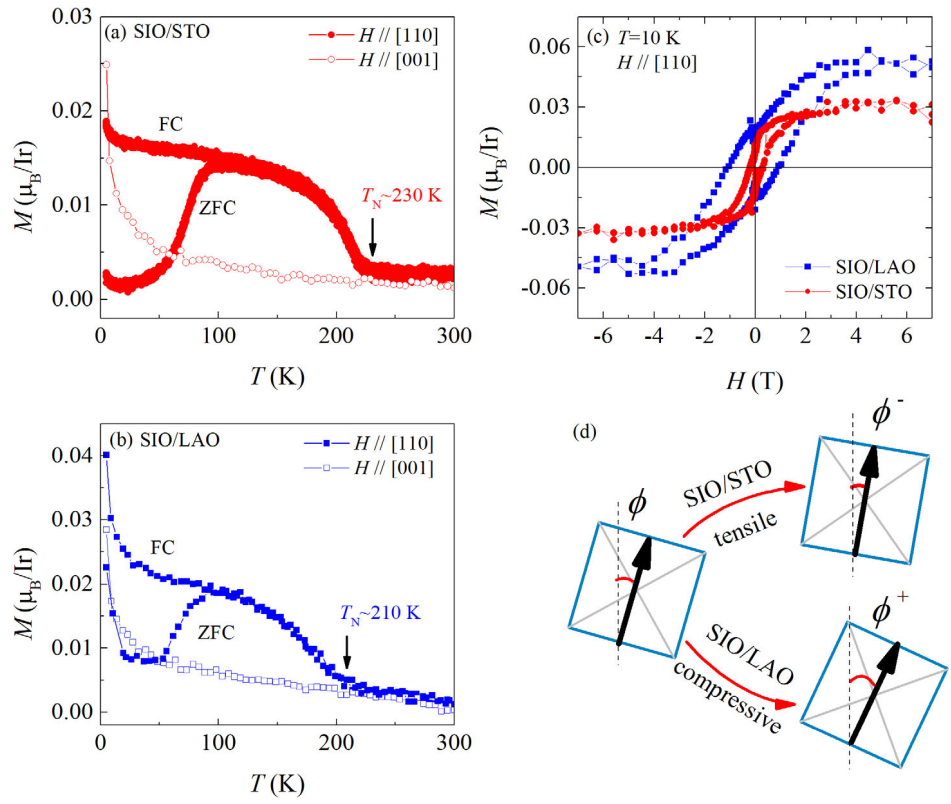

Figure 3

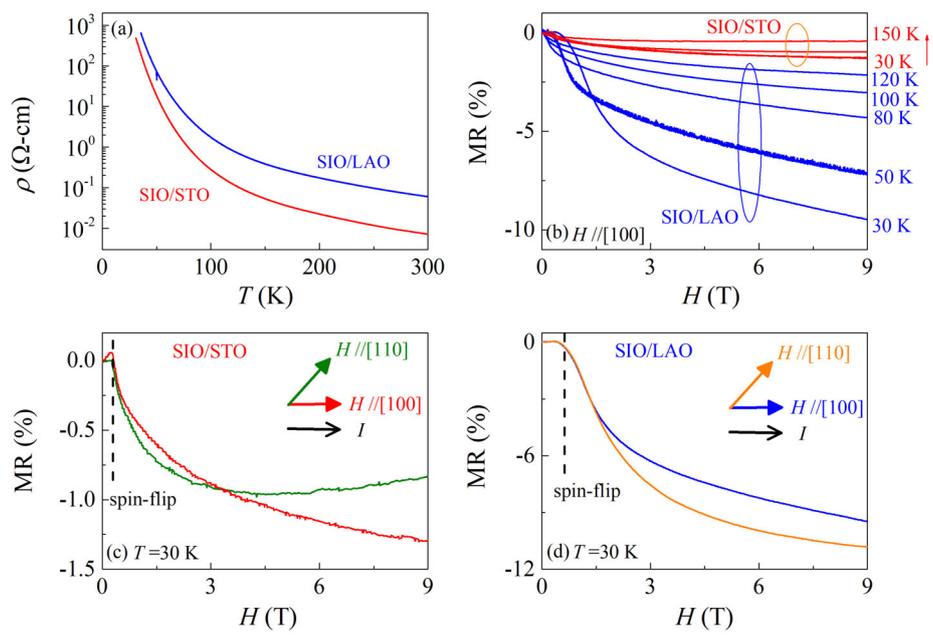

Figure 4

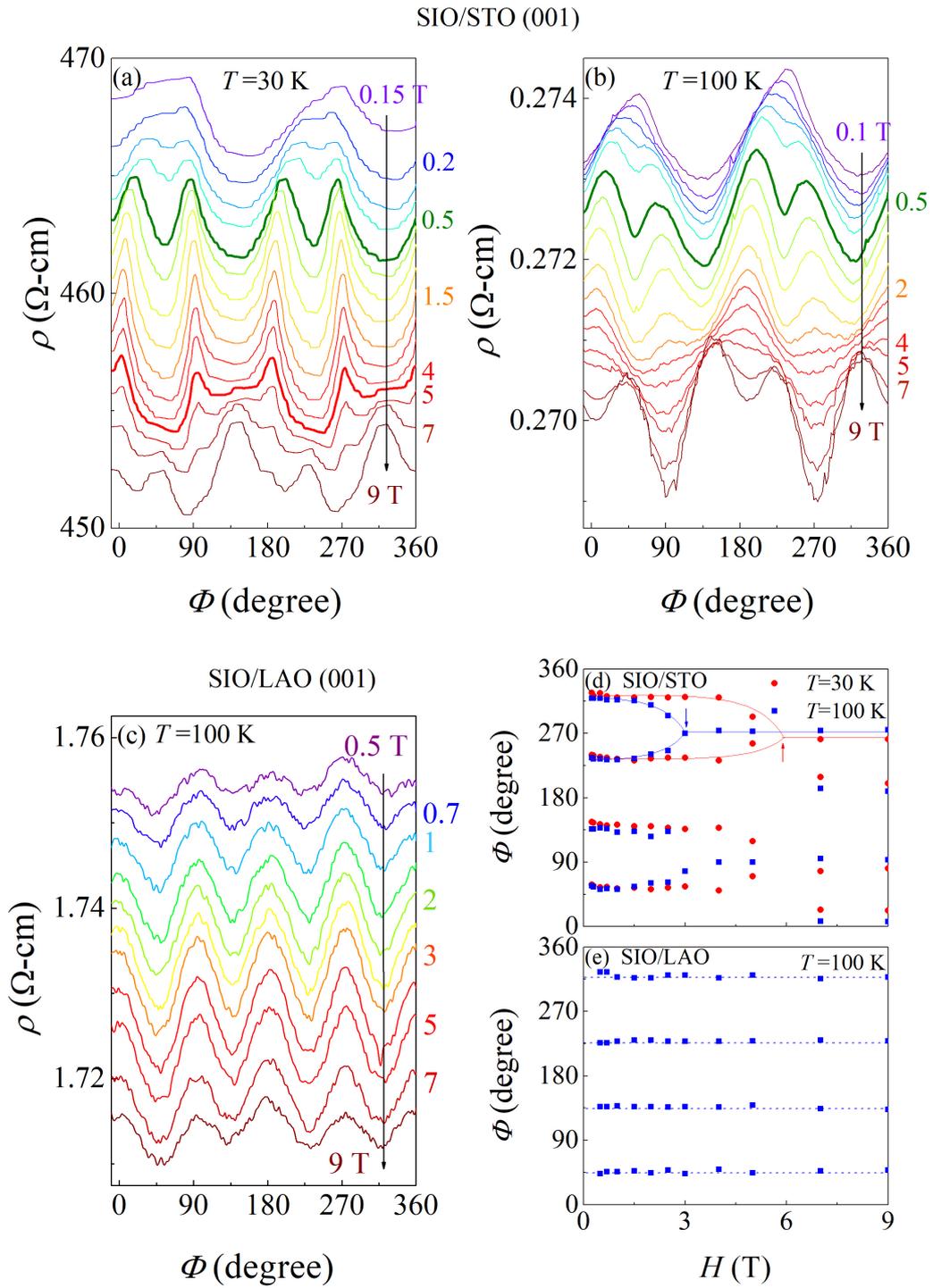

Figure 5